# An introduction to the polaron and bipolaron theoretical concepts

Yuri Kornyushin[a)]
*Maître Jean Brunschvig Research Unit, Chalet Shalva, Randogne, CH-3975*

()

A simple model for the autolocalization of a free charged particle is presented. The polarization well in the model is deep enough for only one localized level. In dielectric materials with a sufficiently large dielectric constant, two charged identical particles can be localized in one polarization potential well, forming a bipolaron. Although several localized levels can be found in more realistic self-consistent models of this type, the more realistic theories require a high level of knowledge of mathematics. Hence, the proposed model can serve as an introduction to the ideas and concepts of autolocalized states.
[DOI: ]

## I. INTRODUCTION

Localized states of current carriers in crystals (polarons) have been known since the 1940s (see Refs. 1 and 2). The concept of localized states plays an important role in many areas of condensed matter physics, including superconductivity.[3] The concept of a polaron was introduced by S. I. Pekar[4] and subsequently developed by many authors.

A polaron is a quasiparticle that interacts with the polarization oscillations of a crystal lattice (especially ionic crystals) such that an autolocalized state of the current carriers arises. An autolocalized state is a particular case of a localized state and is qualitatively different from the free state of a particle. Autolocalization occurs in a homogeneous medium due to the internal properties of the medium. The phenomenon is different from localization in an external potential.

The existing theory of autolocalization[2,3] is a rather developed and successful theory. But its study requires advanced mathematics, including the calculus of variations. Moreover, the theoretical calculations are very tedious, complicated, and must be done numerically. Instead, we will present a simple model for the autolocalization of a free charged particle. This model can be used to introduce the study of autolocalization and help students understand the basic concepts of the subject without performing complicated calculations. The polarization potential well in the proposed model is deep enough for only one localized level. In a more realistic theory several localized levels can be found in a self-consistent potential well.

In dielectric materials with a sufficiently large dielectric constant, two identical charged particles can be localized in one polarization potential well, forming a bipolaron. A bipolaron is also a well known phenomenon[2–4] and is thought to be a carrier of a superconducting current in many materials.[3] The polarization potential well for a bipolaron in the simple model that we will discuss is deep enough for only one localized level, in contrast to more accurate theories in which several levels can exist.[2–4]

When a charged particle is in a localized state, an electrostatic field arises around it. This field polarizes the surrounding medium, so that every small volume of the medium acquires a small dipole moment. These moments are oriented so that there is an attraction between the moments and a particle in a localized state, which means that the interaction energy is negative. So a charged particle in a localized state creates a potential well for itself in which it could be localized if the well is deep enough. If the well is not deep enough, the particle will not be in a localized state.

Let us consider a particle with a wave function $\psi(\mathbf{r})$ and a charge $e$. The charge density is $\rho(\mathbf{r}) = e\psi^*(\mathbf{r})\psi(\mathbf{r})$.[5] Let the corresponding electrostatic potential in the dielectric medium with a dielectric constant $\varepsilon$ be $\varphi_0(\mathbf{r})/\varepsilon$, where $\varphi_0(\mathbf{r})$ is the electrostatic potential of the same particle in a vacuum.[6] The potential of the charged particle in a dielectric medium can be written as

$$\varphi_0(\mathbf{r})/\varepsilon \equiv \varphi_0(\mathbf{r}) - [(\varepsilon - 1)/\varepsilon]\varphi_0(\mathbf{r}). \tag{1}$$

The first term on the right-hand side of Eq. (1) represents the potential of the particle only, and the second term represents the potential produced by the polarization of the dielectric medium. Thus, the second term, $\varphi_d(\mathbf{r}) = -[(\varepsilon - 1)/\varepsilon]\varphi_0(\mathbf{r})$, is the part of the potential that acts on the particle. This potential is attractive because $\varepsilon$ is always greater than unity.

## II. AUTOLOCALIZED CHARGED PARTICLE

To determine the possibility of autolocalization, we assume the following model for the localized wave function and hence for the charge density:

$$\psi(r) = (g^{3/2}/\pi^{3/4})\exp[-0.5(gr)^2], \tag{2a}$$

$$\rho = e(g^3/\pi^{3/2})\exp-(gr)^2. \tag{2b}$$

The quantity $1/g$ in Eq. (2) represents the autolocalization radius, because for $r$ greater than $1/g$, the density of the particle practically vanishes. The model represented by Eq. (2) was chosen because the wave function in Eq. (2a) is similar to those arising in a smooth potential well. Equation (2b) is an exact consequence of Eq. (2a).[5] The total electric field, calculated by using Gauss's theorem, is

$$E(r) = (e/\varepsilon r^2)\Phi(gr) - [2eg/(\pi^{1/2}\varepsilon r)]\exp-(gr)^2, \tag{3}$$

where $\Phi(x)$ is the probability integral, that is, the integral of $[(2/\pi^{1/2})\exp-y^2]$ from $y=0$ to $y=x$. The part of the field that acts on the particle produced by the polarization of the dielectric medium is

$$E_d(r) = [(\varepsilon - 1)/\varepsilon]\{[2eg/(\pi^{1/2}r)][\exp-(gr)^2] - (e/r^2)\Phi(gr)\}. \tag{4}$$

The first term of the expansion in $r$ of the right-hand part of Eq. (3) yields

()

$$E(r) = (4eg^3/3\pi^{1/2}\varepsilon)r + \cdots. \quad (5)$$

From Eq. (5), the electrostatic potential $\varphi(r)$ is

$$\varphi(r) = \varphi(0) - (2/3\pi^{1/2})(eg^3/\varepsilon)r^2 + \cdots, \quad (6)$$

with the part produced by the polarization that acts on the particle given by

$$\varphi(r) = [(\varepsilon-1)/\varepsilon][(2/3\pi^{1/2})eg^3r^2 - \varepsilon\varphi(0)] + \cdots. \quad (7)$$

This approximate potential corresponds to a three-dimensional harmonic oscillator. The model wave function assumed in Eq. (2a) corresponds to the ground state of the three-dimensional harmonic oscillator. The potential represented by Eq. (7) is formed by many electrons and ions. For this reason we can assume that the total mass of all the particles whose charges form the polarization potential [represented by Eq. (7)] is much larger than the mass of the particle under investigation. Hence, it is possible to approximate the reduced mass by the original value of the mass of a particle. This point is a crucial one for the phenomenon of autolocalization. Only the polarization of a medium created by heavy enough particles can lead to the autolocalization of a charged particle, because the structure of the potential, which keeps the particle localized, should be relatively stable with respect to the motion of the localized particle.

An autolocalized current carrier in a crystal together with the surrounding polarization of the medium created by heavy inert particles is usually called a polaron.[1-4] When the motion of a polaron is considered, we have to take into account that the motion of the surrounding cloud of polarization charges accompanies the motion of the localized particle. This phenomenon is expressed by the larger effective mass of a polaron compared to the mass of a bare particle. Hence, the polaron current carriers have low mobility and the sample has high electrical resistance. Now let us calculate the depth of the potential well $\varphi(0)$. For $r \to \infty$, the potential $\varphi(r)$ goes to zero, which means that the value of $\varphi(0)$ is equal to the integral of $E(r)$ from zero to infinity. Hence, we have $\varphi(0) = 2eg/\pi^{1/2}\varepsilon$. The electrostatic potential energy of a charged particle is equal to its charge multiplied by the potential due to the polarized medium, and thus from Eq. (1) it is given by

$$W(0) = -(2/\pi^{1/2})[(\varepsilon-1)/\varepsilon]e^2g. \quad (8)$$

The energy of the particle is, according to Eqs. (7) and (8),

$$W(r) = (2e^2/\pi^{1/2})[(\varepsilon-1)/\varepsilon][(g^3/3)r^2 - g] + \cdots$$
$$= -(2/\pi^{1/2})[(\varepsilon-1)/\varepsilon]e^2g + 0.5m\omega^2r^2 + \cdots, \quad (9)$$

where $m$ is the original effective mass of the particle and $\omega$ is the angular frequency of a harmonic oscillator, which is given by

$$\omega = (2/3^{1/2}\pi^{1/4})[(\varepsilon-1)/\varepsilon]^{1/2}(e/m^{1/2})g^{3/2}. \quad (10)$$

In the ground state the energy of the three-dimensional oscillator is $(\frac{3}{2})\hbar\omega$, and it is smaller than the depth of the potential well $-W(0)$. From this inequality it follows that

$$g < (4/3\pi^{1/2})[(\varepsilon-1)/\varepsilon](me^2/\hbar^2). \quad (11)$$

Equation (11) expresses the condition for the autolocalization of the charged particle in a dielectric medium.

## III. THE EQUILIBRIUM VALUE OF g

The equilibrium value of the inverse autolocalization radius is determined by the minimum of the energy of the system. The system consists of a charged particle and the polarized dielectric around it. Let us consider the intrinsic energy of the polarization itself. The density of the electrostatic energy is a scalar product of **D** and **E** divided by $8\pi$. The part of the induction, **D**, due to the polarization, $\mathbf{D}_p$, is determined by the equation div $\mathbf{D}_p = 0$. Because we are considering the intrinsic energy of the polarization, we should not include the charge of the particle. Because $\mathbf{D}_p$ has a radial component only, we have div $\mathbf{D}_p = (dD_p/dr) + (2/r)D_p = 0$. The only solution of this equation, $D_p = C/r^2$ ($C$ is a constant), goes to infinity at $r=0$ when $C \neq 0$, which is not acceptable. Thus we conclude that $D_p = 0$. This solution corresponds to the absence of a free charge and the presence of a bound charge.[6] The value of the electric field is not zero, but $D_p = E_p + 4\pi P = 0$ ($E_p$ is the electric field due to the polarization, and $P$ is the density of the electric dipole moment of the dielectric, which is formed by the bound charge). Because $D_p = 0$, the intrinsic energy of the polarization is also zero, even though $E_p = -4\pi P \neq 0$.

The intrinsic self-interaction energy of the particle should be excluded. What is left is the energy of the interaction of the particle with the polarized medium induced by the charged particle. The energy at the bottom of the potential well is $W(0)$ [see Eq. (8)], but the particle occupies its ground state with energy $(3/2)\hbar\omega$. Thus, from Eqs. (8)–(10), the interaction energy is

$$W_i = (3^{1/2}/\pi^{1/4})[(\varepsilon-1)/\varepsilon]^{1/2}(e\hbar/m^{1/2})g^{3/2}$$
$$- (2/\pi^{1/2})[(\varepsilon-1)/\varepsilon]e^2g. \quad (12)$$

The right-hand side of Eq. (12) has a minimum at

$$g = g_e = (16/27\pi^{1/2})[(\varepsilon-1)/\varepsilon](me^2/\hbar^2). \quad (13)$$

For the typical value $\varepsilon = 5$, $g_e = 0.267(me^2/\hbar^2)$.

At $g = g_e$ the inequality (11) yields $(\frac{4}{9}) < 1$. The ratio of $(\frac{3}{2})\hbar\omega$ to $-W(0)$ at $g = g_e$ is equal to $\frac{2}{3}$, which means that at least one level (the ground state level) could be formed in the potential well. But the energy of the first excited level at $g = g_e$ is $(\frac{3}{2})\hbar\omega + \hbar\omega = 2.5\hbar\omega$, which is larger than the depth of the potential well $-W(0)$ (the ratio of $2.5\hbar\omega$ to $-W(0)$ is $\frac{10}{9}$). Hence, the potential well in our model is not deep enough to have a first excited level. Moreover, the approximate model potential (which is derived from the model wave function) is not correct for the wave function of the first excited state of a three-dimensional oscillator. In a more realistic theory several localized levels often can be found in a self-consistent potential well.[2-4] So our model is not suitable for the description of such a case.

## IV. THE BINDING ENERGY

The binding energy is determined from Eqs. (12) and (13):

$$W_b = W_i(g_e) = -(32/81\pi)[(\varepsilon-1)/\varepsilon]^2(me^4/\hbar^2). \quad (14)$$

For $\varepsilon = 5$, we have $W_b = -0.0805(me^4/\hbar^2) = -2.19$ eV, which corresponds to a temperature of 25415 K, which is much larger than the ambient temperature. Because the depth of the potential well is much larger than $kT$, all the electrons are in an autolocalized state. So in an autolocalized state, the electrons in a conductivity band have considerably lower energy than in a free (plane wave) state. This lower energy influences the concentration of the electrons in a conductivity band, because their concentration depends exponentially on the gap energy.[7] (The gap energy is the difference between the electron energy at the bottom of the conductivity band and at the top of the valence band.) The concentration of the current carriers significantly influences the electrical conductivity and optical properties. Optical properties are in a great extent dependent on the plasma frequency,[7] which is proportional to $(n/m_a)^{1/2}$ ($n$ is the number density of the current carriers in a conductivity band and $m_a$ is the effective mass of an electron in an autolocalized state). So autolocalization significantly affects various physical properties.

## V. CURRENT CARRIERS (ELECTRONS AND HOLES) IN A SEMICONDUCTOR

The present approach is valid when it is possible to consider a separate localized current carrier and to neglect the presence of the other current carriers. This assumption is valid when the Debye screening radius $1/g_D$ is much larger than the localization radius of the current carrier $1/g_e$. When the current carriers present in the sample are of one type only and are not degenerate,[7] which means that the density of the current carriers is not too high and the temperature is not too low, then

$$g_D^2 = 4\pi e^2 n/\varepsilon kT. \quad (15)$$

We have from Eqs. (13) and (15) the inequality

$$n \ll n_u = 0.00890[(\varepsilon-1)^2/\varepsilon]kT(m^2e^2/\hbar^4). \quad (16)$$

Here $n_u$ is the upper limit of the concentration of the current carriers, defined by Eq. (16). On the other hand, to consider the current carriers as noninteracting, we should find much less (on the average) than one current carrier inside the sphere of a Debye screening radius. In this case every current carrier is screened from the influence of the others. Thus $4\pi n/3q_D^3 \ll 1$, and it follows that

$$n \gg n_l = (1/36\pi)(\varepsilon kT/e^2)^3. \quad (17)$$

Here $n_l$ is the lower limit of the concentration of the current carriers, defined by Eq. (17). The inequalities (16) and (17) can be combined to give

$$n_l \ll n \ll n_u. \quad (18)$$

For $m$ equal to the electron mass, $\varepsilon = 5$, and $T = 300$ K, we have $n_l = 6.41 \times 10^{15}$ cm$^{-3}$ and $n_u = 1.82 \times 10^{20}$ cm$^{-3}$. Because the concentration of the current carriers in semiconductors is about $10^{17}$–$10^{18}$ cm$^{-3}$,[6] the inequalities in Eq. (18) are satisfied by the properties of many materials.

## VI. BIPOLARON

Let us consider two identical particles in a potential well. Two current carriers in a self-consistent potential well are called a bipolaron.[2–4] Bipolarons play an important role in solid state physics, and they can be superconducting current carriers in some materials.[3] Both of the charged particles of a bipolaron occupy the ground state (they could be bosons or fermions with opposite spins in general). Let $\psi$, $\rho$, $E(r)$, $E_0(r)$, and $E_d(r)$ be the same as before for each particle. It follows from Eq. (9) that the potential for the two particles, each one with charge $e$ (the total charge is $2e$), is

$$W(r) = -(4/\pi^{1/2})[(\varepsilon-1)/\varepsilon]e^2 g$$
$$+ (4/3\pi^{1/2})[(\varepsilon-1)/\varepsilon]e^2 g^3 r^2 + \cdots. \quad (19)$$

The last term on the right-hand side of Eq. (19) is equal to $0.5 m\omega_b^2 r^2$. It follows that

$$\omega_b = (2^{3/2}/3^{1/2}\pi^{1/4})[(\varepsilon-1)/\varepsilon]^{1/2}(e/m^{1/2})g^{3/2}. \quad (20)$$

The direct interaction energy of the identical particles should be taken into account. It is determined by the integral of $2E_0^2(r)$ over all space divided by $8\pi$. This integral yields

$$W_i = (2/\pi)^{1/2}e^2 g. \quad (21)$$

The energy of the two particles in the ground state of a potential well is $3\hbar\omega_b$. In an autolocalized state the sum of this energy and $W_i$ is smaller than the depth of the potential well $(4/\pi^{1/2})[(\varepsilon-1)/\varepsilon]e^2 g$ [see Eq. (19)]. It follows that

$$0 < g^{1/2} < (1/2^{3/2} 3^{1/2}\pi^{1/4})[(\varepsilon-1)/\varepsilon]^{1/2}\{[(4-2^{1/2})\varepsilon-4]/$$
$$(\varepsilon-1)\}(m^{1/2}e/\hbar). \quad (22)$$

Equation (22) is the condition for the existence of at least one localized level in a self-consistent potential well. Note that $g^{1/2}$ is positive when $\varepsilon > 4/(4-2^{1/2}) = 1.55$ only. When $g^{1/2}$ is negative, there is no localization because the wave function is not localized. The direct interaction of the two particles results in an increase of the energy of the system. The attractive interaction of the particles with the polarization of the medium leads to a decrease of the energy. But if $\varepsilon$ is not big enough, this decrease is not sufficiently large to compensate for the energy increase due to the repulsion of two particles. In this case the resulting potential well is not deep enough and the ground state cannot be formed. In such materials a bipolaron does not exist.

The total energy of the system of two particles follows from Eqs. (19)–(21) and is given by

$$W(g) = (2^{3/2} 3^{1/2}/\pi^{1/4})[(\varepsilon-1)/\varepsilon]^{1/2}(e\hbar/m^{1/2})g^{3/2}$$
$$+ (2/\pi)^{1/2}e^2 g - (4/\pi^{1/2})[(\varepsilon-1)/\varepsilon]e^2 g. \quad (23)$$

The minimum of this energy corresponds to

$$g_e^{1/2} = (1/2^{1/2} 3^{3/2}\pi^{1/4})[(\varepsilon-1)/\varepsilon]^{1/2}\{[(4-2^{1/2})\varepsilon-4]/\varepsilon\}$$
$$\times (m^{1/2}e/\hbar). \quad (24)$$

We can see that $g_e^{1/2}$ is positive for $\varepsilon > 4/(4-2^{1/2}) = 1.55$, in accord with the previous result. Equation (24) can be rewritten as

$$g_e = (1/54\pi^{1/2})[(\varepsilon-1)/\varepsilon]\{[(4-2^{1/2})\varepsilon-4]/\varepsilon\}^2(me^2/\hbar^2). \quad (25)$$

For $\varepsilon = 5$, Eq. (25) yields $g_e = 0.0416(me^2/\hbar^2)$. This value of the inverse of the autolocalization radius is much smaller than that calculated from Eq. (13), which means that the localization radius of a bipolaron is much larger than that of an autolocalized single charged particle.

The equilibrium energy of the system according to Eq. (23) is

$$W(g_e) = -(1/162\pi)\{[(4-2^{12})\varepsilon-4]^3/$$
$$\varepsilon^2(\varepsilon-1)\}(me^4/\hbar^2). \quad (26)$$

Again the minimum energy $W(g_e)$ is positive for $\varepsilon > 4/(4-2^{1/2}) = 1.55$, in accord with our previous result. For $\varepsilon = 5$, the minimum energy $W(g_e) = -0.0140(me^4/\hbar^2)$, which is much smaller than the binding energy of two separate particles. The autolocalized state of two particles (bipolaron) is a metastable state because its minimum energy is higher than that of the two separate particles.

## VII. DISCUSSION

We have proposed a model that describes the concept of autolocalization. The advantage of this model is that it allows us to study the basic concepts of autolocalization without using complicated and mainly numerical calculations.

We have shown that localized states can form in a regular dielectric or semiconductor with a sufficient (but not too high) density of current carriers. Only one (ground) level can be formed in a polarization potential well. If the dielectric constant is greater than 1.55, bipolaron states could exist as metastable states, because the binding energy of two separate current carriers is larger than the binding energy of the two carriers in one bipolaron well. Thus, it is obvious that there exists only one state in our model of a bipolaron polarization well. The localization radius of a bipolaron is considerably larger than that of a polaron.

Of course, reality is much more complicated than the model discussed here. Several excited levels of a polaron and bipolaron can exist, which also influence the properties of solids.[3,4]

## VIII. SUGGESTED PROBLEM

Consider a charged particle in a vacuum. Calculate and analyze the part of the electrostatic potential that acts on the particle.

Solution: It is well known in quantum electrodynamics that near the point charge in a vacuum, the electrostatic potential has the form[8]

$$\varphi(r) = (e/r) - (2\alpha e/3\pi r)[1.41 - \log(\hbar/mcr)], \quad (27)$$

where $\alpha$ is a constant, $m$ is the (bare) mass of the particle,[8] and $c$ is the speed of light in vacuum. Equation (27) is valid for $r$ smaller than $\hbar/mc$.[8] The first term in Eq. (27) describes the part of the potential produced by the bare charged particle, while the second term describes the part of the potential, produced by the polarization of the vacuum.

Because there is no self-interaction in nature, the part of the electrostatic potential that acts on the particle is [see Eq. (27)]

$$\varphi_a(r) = -(2\alpha e/3\pi r)[1.41 - \log(\hbar/mcr)]. \quad (28)$$

To analyze this potential we introduce the notation $x = \hbar/mcr$. Then

$$\varphi_a(x) = (2\alpha emc/3\pi\hbar)[(x\log x) - 1.41x]. \quad (29)$$

This function has a minimum at $x = x_e = \exp(0.41) = 1.51$, which corresponds to $r_e = 0.664(\hbar/mc)$. The minimum energy is $e\varphi_e = -3.01\alpha e^2 mc/3\pi\hbar$. At $\alpha = 1$ for an electron, $e\varphi_e = -1191$ eV. It looks like that at such a deep energy minimum the wave function of the particle can be localized within the sphere of radius $r_e$. But the potential of the well, $\varphi_a(r)$, in the vicinity of the energy minimum is almost zero for $r > r_e$. So the well is very narrow, kinetic energy is very high and the localization at $r \approx r_e$ is impossible as expected.